\documentclass[
aps,%
12pt,%
final,%
notitlepage,%
oneside,%
onecolumn,%
nobibnotes,%
nofootinbib,%
superscriptaddress,%
noshowpacs,%
centertags]%
{revtex4}
\usepackage{graphicx,times}
\usepackage{longtable}
\usepackage{amssymb,amsmath}
\usepackage{caption2}
\usepackage{xcolor}
\usepackage{multirow}

\usepackage[T2A]{fontenc}
\usepackage[utf8]{inputenc}
\usepackage[english]{babel}

\usepackage{amssymb,amsmath,amsthm}
\usepackage{systeme,mathtools}
\makeatletter
\renewcommand*\env@matrix[1][*\c@MaxMatrixCols c]{%
  \hskip -\arraycolsep
  \let\@ifnextchar\new@ifnextchar
  \array{#1}}
\makeatother
\usepackage{lipsum}
\usepackage{relsize}

\usepackage[normalem]{ulem}

\def\squareforqed{\hbox{\rlap{$\sqcap$}$\sqcup$}}

\def\sq{\ifmmode\squareforqed\else{\unskip\nobreak\hfil
\penalty50\hskip1em\null\nobreak\hfil\squareforqed
\parfillskip=0pt\finalhyphendemerits=0\endgraf}\fi}

\def\arcmin{\hbox{$^\prime$}}

\def\utw{\smash{\rlap{\lower5pt\hbox{$\sim$}}}}

\def\udtw{\smash{\rlap{\lower6pt\hbox{$\approx$}}}}

\def\diameter{{\ifmmode\mathchoice
{\ooalign{\hfil\hbox{$\displaystyle/$}\hfil\crcr
{\hbox{$\displaystyle\mathchar"20D$}}}}
{\ooalign{\hfil\hbox{$\textstyle/$}\hfil\crcr
{\hbox{$\textstyle\mathchar"20D$}}}}
{\ooalign{\hfil\hbox{$\scriptstyle/$}\hfil\crcr
{\hbox{$\scriptstyle\mathchar"20D$}}}}
{\ooalign{\hfil\hbox{$\scriptscriptstyle/$}\hfil\crcr
{\hbox{$\scriptscriptstyle\mathchar"20D$}}}}
\else{\ooalign{\hfil/\hfil\crcr\mathhexbox20D}}%
\fi}}

\newcommand{\aaa}{Astronom. and Astrophys.}
\newcommand{\aas}{Astronom. and Astrophys. Suppl. Ser.}

\begin{document}
\selectlanguage{english}

\title{Search for Galaxy Cluster Candidates in the Cosmic Microwave Background Maps of the Planck Space Mission Using a Convolutional Neural Network Based on the Method of Tracing the Sunyaev–Zeldovich Effect}

\author{O. V. Verkhodanov}
\affiliation{Special Astrophysical Observatory, Russian Academy of Sciences, Nizhnii Arkhyz, 369167 Russia}

\author{A. P. Topchieva}
\email{ATopchieva@inasan.ru}
\affiliation{Institute of Astronomy, Russian Academy of Sciences, Moscow, 119017 Russia}

\author{A. D. Oronovskaya}
\affiliation{Astrophysical School ``Traektoria'', Moscow, 109147 Russia}

\author{S. A. Bazrov}
\affiliation{Astrophysical School ``Traektoria'', Moscow, 109147 Russia}

\author{D. A. Shorin}
\affiliation{Astrophysical School ``Traektoria'', Moscow, 109147 Russia}


\begin{abstract}
We propose a method of searching for radio sources exhibiting the Sunyaev–Zeldovich effect in the multi-frequency emission maps from the Planck mission data using a convolutional neural network. A catalog for recognizing radio sources is compiled using the GLESP pixelation scheme at the frequencies of 100, 143, 217, 353, and 545 GHz. The quality of the proposed approach is evaluated and the quality of the dependence of model data on the $\text{S}/\text{N}$ ratio is estimated. We show that the presented neural network approach allows the detection of sources with the Sunyaev–Zeldovich effect. The proposed method can be used to find the most likely galaxy cluster candidates at large redshifts.


ISSN 1990-3413, Astrophysical Bulletin, 2021, Vol. 76, No. 2, pp. 124–132. c Pleiades Publishing, Ltd., 2021.
\end{abstract}

\maketitle
\section{INTRODUCTION}
\label{sec:intro}

One of the results of the work of the Planck space telescope is the creation of a cosmic microwave background intensity distribution for the entire sky. Among other data, this map includes sources exhibiting the Sunyaev--Zeldovich effect (hereafter SZ---effect), which can be used to identify galaxy clusters. The SZ--effect is a result of cosmic microwave background photon scattering on hot plasma electrons. Conditions suitable for observations are exhibited by galaxy clusters~\cite{zs}. Studying such clusters at different redshifts helps in our understanding of the working principles of the Universe by means of verifying cosmological models and limiting the cosmological constants~\cite{Allen, Kravtsov, Nagai}. Such objects are also used in solving astrophysical problems~\cite{Sarazin, Kormendy, Bykov}. Studying galaxy clusters in millimeter and submillimeter range, observable due to the SZ--effect \cite{zs}, in the X--ray range where hot gas radiates, as well as in the visual remains a fundamental direction in cosmology. Thus, we can trace the cluster mass evolution and especially the formation of large-scale structures in the Universe in different cosmological epochs~\cite{Basu}.

Observations of the effect are complicated due to its small amplitude, experimental data errors, and the distortion of the cosmic microwave background temperature by other sources. Since the SZ--effect is a scattering effect and its magnitude does not depend on redshift, clusters at high redshifts may be detected as simply as those at small redshifts, if their mass is sufficient ($M_{\rm clust}\sim10^{14} M_{\rm sun} $, where $M_{\rm clust} = M_{\rm 500}$). The angular size to redshift ratio is also a factor that helps detect clusters at high redshifts: it varies little between the redshifts 0.3 and 2, which implies that clusters with redshifts in the stated range have practically identical sizes in the sky.

Earlier we already used the approach based on selecting cluster candidates using the SZ--effect in the direction of radio sources (see \cite{rc_planck,wenss_sz,traj_zaporozhets}). In this study we continue developing the field related to an automated SZ-object selection algorithm based on machine learning methods (see the description of the first experiment in \cite{oron_habr}).

Methods of deep machine learning are now used more and more often in astronomical studies. To name a few, let us mention studies on dust distribution and its properties in galaxies \cite{Dobbels}, the search for connections between stars in the centers of galaxies and the morphology of those galaxies (see ~\cite{Tacchella}), determining galactic parameters such as density, metallicity, surface brightness and ionization degree from galactic spectra emission lines in the visual, ultraviolet, infrared and submillimeter ranges (see~\cite{Ucci}). The methods of machine learning were also used to select 400 galaxies with high jet activity indices in the Sloan Digital Sky Survey spectra of two million galaxies \cite{Baron}. Similar algorithms are also used in the classification of extended sources in the radio range, in studies of radio galaxies based on their morphological properties using convolutional neural networks \cite{Aniyan}, and in spectral classification of galaxies \cite{Tao}. Finally, note also the creation of a system of machine learning while searching for the SZ--effect using Planck HFI data \cite{Bonjean2019}; 18 thousand SZ--object candidates were selected from the high-frequency data as a result.

In this study we present a machine learning algorithm that can help expand the list of galaxy clusters in the millimeter range. There are observational data available on the microwave background from Planck \cite{planck_zs}, SPT \cite{spt_zs}, and other observatories, e.g., ACT \cite{act_zs}, which are basically sky maps at various wavelengths. The SZ--effect can be detected in some regions of the map (in the direction of the cluster) and has a special spectral shape at 217 GHz and below the signal is lower than the microwave background, while at higher frequencies it increases. We aim to find such regions on the map. Unlike the methods developed earlier \cite{Bonjean2019, Melin, Herranz}, in this paper we analyze the completeness of the obtained sample by comparing with simple Matched multi-filter (MMF1 \cite{Herranz}, MMF3 \cite{Melin}) and PowellSnakes (PwS \cite{Bonjean2019}) algorithms for finding objects. We also estimate the quality of the created model and the effect of $\text{S}/\text{N}$ ratio variations on the results of network learning, which may be useful when working with lower quality data. Section 2 describes the data format used for training the neural network, as well as methods used in obtaining and reduction. Section 3 presents the algorithm itself and examples of its application to the Planck observatory maps. Section 4 presents the results obtained by the neural network, the ResNet18 and RandomForest machine learning model comparison with each other and also with the MMF1, MMF3, and PwS methods based on Recall metrics using the ResNet18 approach, as well as the detection quality of the considered objects using model data for arbitrary Gaussian maps with various $\text{S}/\text{N}$ ratios. Finally, we formulate the Conclusions for the presented study. The metrics and approach algorithm for analyzing sources with the SZ--effect are described in the Appendix.

\section{DATA REDUCTION}
\label{sec:data}
This work uses data taken from the Planck Legacy Archive~\footnote{\tt https://pla.esac.esa.int/\#home} . These data were converted to GLESP format. The GLESP (Gauss–Legendre Sky Pixelization) ~\cite{glesp1} software package uses cosmic microwave background (CMB) map pixelization schemes, based on using Gauss–Legendre polynomial zeros to organize the sky division grid, and allows to obtain a strictly orthogonal map layout. The HEALPIX package is widely used when working with the CMB data, however, we used the GLESP software package since the authors are familiar with it. For the purpose of of this study, both packages are identical in terms of their capabilities. The package contains procedures for working with individual images in areas of fixed size and allows to quickly convert temperatures in coordinates to a temperature spectrum, i.e., to expand in spherical functions using a special Gauss integration technique. To create a sample of objects with SZ-effect , we selected the maps with a priori present SZ--effect without the SZ--effect were chosen by random coordinates in the region that is known to have no SZ--effect. The capabilities of GLESP are described in more detail in \cite{glesp1,glesp3}. The data on the derived sample of sources can be found in the open access SAO RAS archive \footnote{\tt http://sed.sao.ru/vo/planck\_maps/}.

The sample of SZ--effect sources was selected from three catalogs: ABELL (Abell Clusters) \cite{Abell}, PSZ1 \cite{Planck1}, and the second Planck data archive PSZ2 \cite{Planck2}. Regions with no SZ-effect were chosen by eye in random fields. We checked for the presence of objects in the maps used as no SZ--effect maps. The object absence criterion was a lack of signal at the 217 GHz frequency and higher. This was done prior to training the model, in order to avoid adding incorrect data to the sample which is important when the number of objects is small. For a larger sample this verification can be omitted. In order to train the network, we created a catalog of images with SZ-effect, which consisted of $1 000$ sources with data at 100, 143, 217, 353, and 545\,GHz.

Due to the typical SZ--effect characteristic --- the signal intensity decrease at 217 GHz and lower --- we can identify galaxy cluster candidates and isolate them from the signals of other radio sources, which do not exhibit this effect \cite{Planck2}.

In addition to the GLESP being used for converting the Planck telescope observations, it was also used to cut out $30\arcmin\times30\arcmin$ regions in the vicinities of objects with SZ-effect at 100, 143, 217, 353, and 545\,GHz.

Thus obtained dataset consists of two classes: images with and without SZ--effect. Each class contains 1000 objects, each object consists of five images.

\begin{figure}
a) 100 \includegraphics[width=0.14\linewidth]{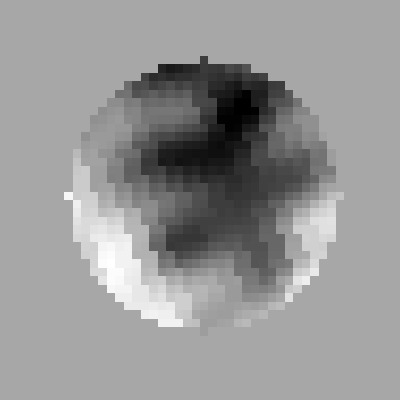}
143 \includegraphics[width=0.14\linewidth]{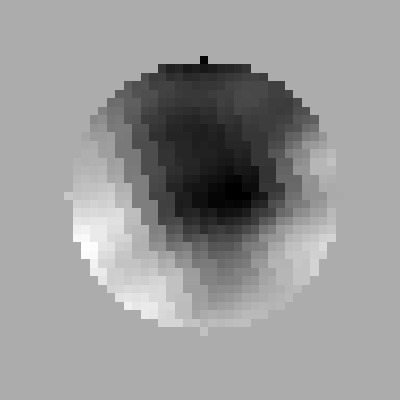}
217 \includegraphics[width=0.14\linewidth]{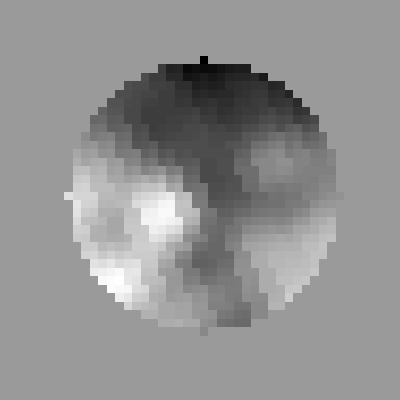}
353 \includegraphics[width=0.14\linewidth]{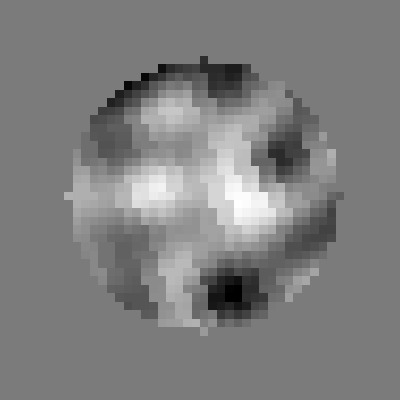}
545 \includegraphics[width=0.14\linewidth]{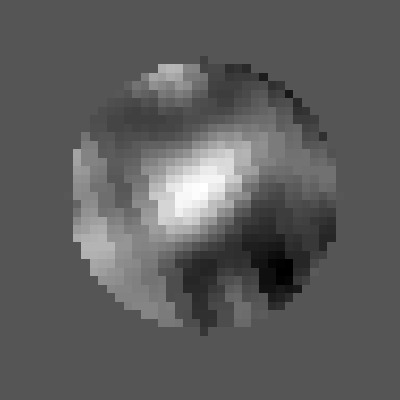}\\
b) 100 \includegraphics[width=0.14\linewidth]{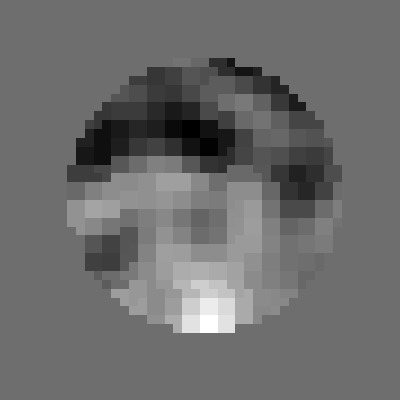}
143 \includegraphics[width=0.14\linewidth]{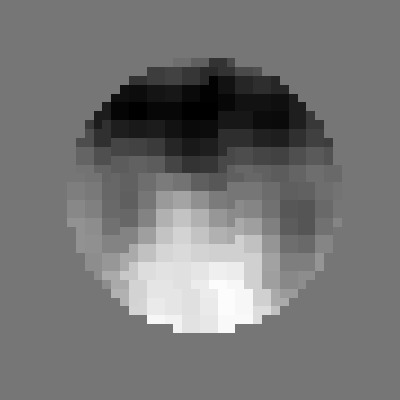}
217 \includegraphics[width=0.14\linewidth]{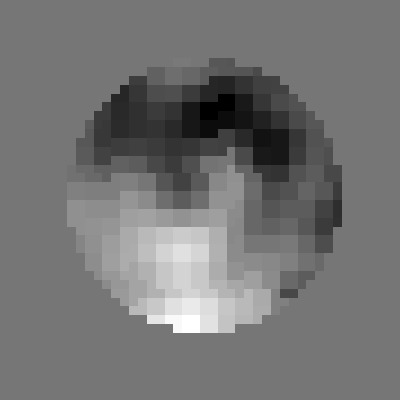}
353 \includegraphics[width=0.14\linewidth]{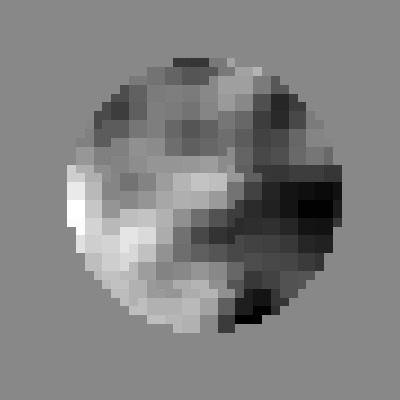}
545 \includegraphics[width=0.14\linewidth]{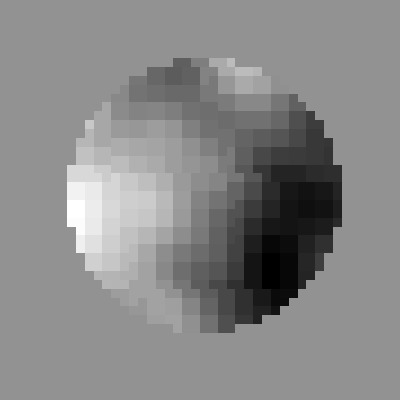}
\caption{An example of sample objects used for neural network learning at 100, 143, 217, 353, and 545\,GHz correspondingly: (a)---object PSZ2 G008.94-81.22 with SZ-effect; (b)---sky region with no SZ-effect.}
\label{fig:freq11}
\end{figure}

For each source we prepare images of $30\arcmin$ sky regions. When transferring the image to a new pixel grid, we assign to each grid node the value of the closest node in the old grid (the so-called ``nearest neighbor interpolation'' method). The new grid is needed to decrease the size of the images, since much machine resources are used when using the initial images. Learning by large-sized images in not feasible, since the increased image does not carry any additional information. We used a circular mask around both types of objects. Fig.~\ref{fig:freq11} shows an example of the object PSZ2 G008.94-81.22 from the Planck catalog \cite{Planck1} with the SZ-effect.

We divided the sample into training, validating, and testing samples in a ratio of 70/15/15, putting 150 random objects from each class into the validation sample and 150 into the test sample. Thus, all three derived samples are balanced. To expand the training sample, we used augmentations (small transformations) of initial objects: a turn by a random angle and slight horizontal and vertical shifts. It is important that the augmentations were the same at all frequencies for the same object.

\section{DESCRIPTION OF THE ANALYTICAL ALGORITHM}
\label{sec:algorithm}

We are solving the problem of binary classification. One of the first classification neural networks was the AlexNet~\cite{AlexNET1}, which was developed into the VGG~\cite{VGG1} architecture. Both architectures use blocks of convolutional layers and pooling, i.e. a gradual decrease of spatial resolution. The last mini-batch normalization layer was reset to zero, and each consecutive residual branch has several fully connected layers. Authors~\cite{AlexNET1, VGG1} nominally specify that the convolutional blocks are responsible for the extraction of patterns, whereas the aim of the fully connected layers is to combine the extracted patterns to obtain the solution. These architectures have several limitations. First, the fully-connected layers contain many parameters and are the most computationally demanding part of the models. Second, deeper architectures based on such models do not learn due to gradient attenuation~\cite{Hochreiter}.

These problems were solved by the architectures ResNet~\cite{resNET1}and InceptionNet~\cite{InceptionNet1} in the open access PyTorch machine learning library~\footnote{\tt https://github.com/pytorch/vision/blob/master/torchvision/models/resnet.py}. ResNet uses ``residual connections'' in each convolutional block, which allows the gradient to flow through them without being attenuated. A learning parameter is present in each block which regulates the fraction of the signal that passes through the residual connection. The network has a possibility of converging to a solution where the entire signal will pass through a residual connection in part of the blocks, i.e. some blocks will be turned off. Thanks to the ``residual connections'' it was possible to train very deep ResNet-models (up to 152 layers)~\cite{resNET1}. InceptionNet uses intermediate exits from the model, which also helps to avoid gradient attenuation during training. Further development of the architectures was divided into two directions: models that give the best classification quality, and models which are computed rapidly (for example, MobileNet~\cite{MobileNet}).

Due to the small amount of data in the training set we cannot use deep architectures aimed at reaching maximum quality through laborious computations. We trained MobileNet, VGG, and finally settled to ResNet18. This network shows the quality comparable to the other models, but at the same time gives a greater quality metrics stability for various sample divisions into training and testing parts. We assumed that such an architecture is more suitable for this goal. Since the description of this model architecture is presented in detail in the developers’ paper, we shall omit it here. We also omit a detailed result comparison for different architectures, since our main goal is to study the fundamental possibility of using the neural network approach in detecting objects.

The PyTorch~\cite{NEURIPS2019_9015} library was used as a software tool for training. Our network accepts fivechannel images, which makes it impractical to transfer the weights from a network trained by three-channel (RGB) images from ImageNet \cite{Russakovsky}. Binary cross-entropy is used as the Loss function. The neural network trains using the stochastic gradient descend by the RAdam \cite{liu2019radam} optimizer. We took standard RAdam parameter values, with the exception of $\text{learning rate} = 2\cdot10^{-4}$. This parameter was selected based on the validation sample.

\section{RESULTS}
\label{sec:results}
\subsection{Comparison of ResNet18 and RandomForest}

To estimate the quality of work classification we used such metrics as Accuracy, Recall, Precision, F1, ROC AUC. The Accuracy metric characterizes the fraction of correctly classified objects. The use of Accuracy is justified, since the classes in the presented dataset are balanced. The Recall characterizes the fraction of correctly identified objects out of those presented in the sample, and the Precision—the fraction of correctly identified objects out of the predicted ones. The F1-measure links the Precision and Recall metrics and is introduced as their harmonic mean~\citep{Fmera}. If F1 is close to unity, the classifier model is the most accurate. ROC (Receiver Operating Characteristic) is a graphic binary classifier quality characteristic, it is the dependence of the fraction of true positive classifications on the fraction of false positive classifications when varying the threshold of the deciding rule. The ROC AUC metric is the area below the ROC curve and is often used to estimate the quality of separating objects of two classes by this algorithm. The metrics properties are described in more detail in the Appendix.

The neural network was trained on the training set during 100 epochs (the "epoch" parameter characterizes the level of training of the network: the number of runs through the entire training set), after which a comparison was made by the described metrics. We selected only the model weights for which the Loss function was minimal in the validation set.

For comparison, we also tested the possibility of classifying using a method that does not involve neural networks. In this case, the features should be prepared manually. For each channel, we selected the ``mean'', ``dispersion'', and also a five-bin histogram with fixed thresholds as features. We used the ``tree ensemble'' model (RandomForest) with the number of trees --- 300, number of objects --- 1000, minimal number of objects in a node --- 2, number of variables in the set --- 10. The aim of this experiment was to show that the task cannot be solved ``directly'', and using neural networks for searching for objects with SZ--effect is justified. The comparison results are presented in Table~\ref{table:comparison}.

\begin{table}[!h]
\onelinecaptionstrue
\setcaptionmargin{5mm}
\begin{tabular}{|l|l|l|l|l|}
\hline
Алгоритм & F1 & Accuracy & Recall & ROC AUC \\
\hline
MMF1 & - & - & 0.747 & - \\
MMF3 & - & - & 0.740 & - \\
PwS & - & - & 0.620 & - \\
RandomForest & \textbf{0.646} & \textbf{0.653} & \textbf{0.633} & \textbf{0.716} \\
ResNet18 & \textbf{0.891} & \textbf{0.893} & \textbf{0.873} & \textbf{0.946} \\
\hline
\end{tabular}
\caption{Object classification results for the test sample. ResNet18 and RandomForest are the methods used in this work.}
\label{table:comparison}
\end{table}

\begin{figure}
\includegraphics[width=0.8\linewidth]{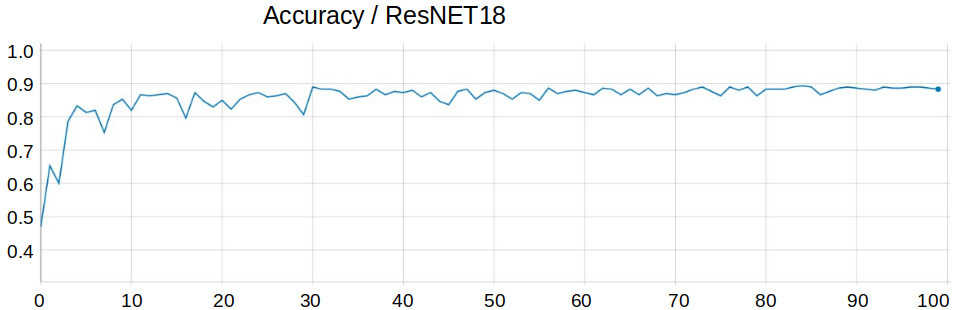}
\caption{Variation of the numerical value of the Accuracy metric for the validation data as a function of epoch.}
\label{fig:freq1122}
\end{figure}

\begin{figure}
\includegraphics[width=0.8\linewidth]{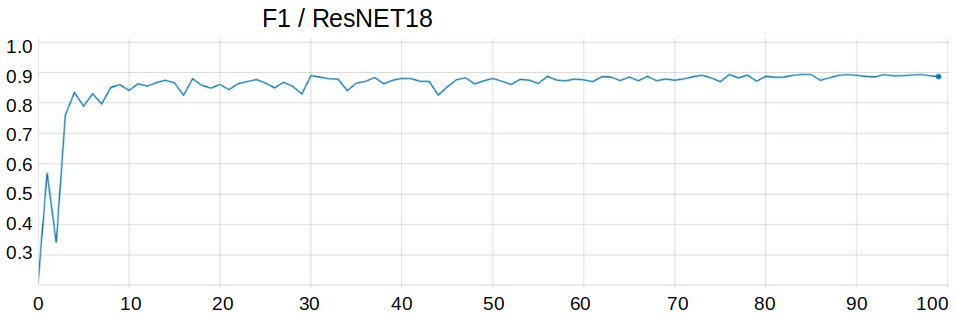}
\caption{Variation of the numerical value of F1 for the validation data as a function of epoch.}
\label{fig:freq1123}
\end{figure}

According to the results presented in Table~\ref{table:comparison}, ResNet18 outperforms RandomForest significantly by each of the presented metrics, which shows the efficiency of the proposed approach. The advantage of the approach using ResNet18 is that there is no need to prepare the patterns for learning, the network accepts features the images themselves at five frequencies. The same is true for estimates and comparison of the method by Recall: ResNet18 also outperforms RandomForest. Fig.~\ref{fig:freq1122}  shows the change in the Accuracy metric during training, which characterizes the fraction of correctly classified objects. We can conclude that the fraction of correctly classified objects increases during the learning. Fig.~\ref{fig:freq1123} shows how the numerical value of the F-measure increases during the learning, therefore, the quality of the model improves.

\begin{figure}
\includegraphics[width=0.8\linewidth]{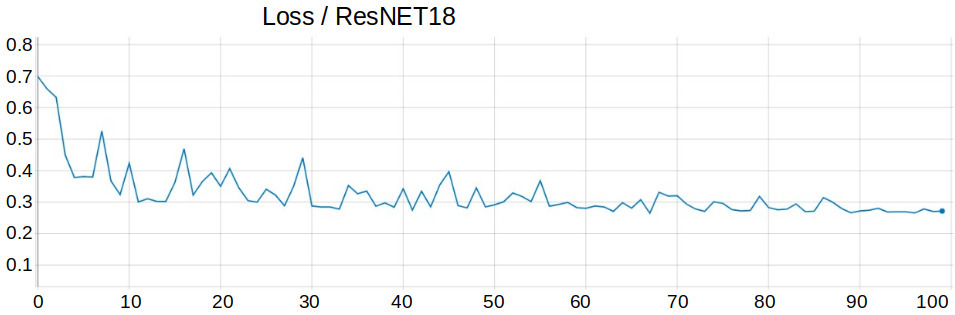}
\caption{Variation of the numerical value of Loss for the validation data as a function of epoch.}
\label{fig:freq1124}
\end{figure}

\begin{figure}
\includegraphics[width=0.8\linewidth]{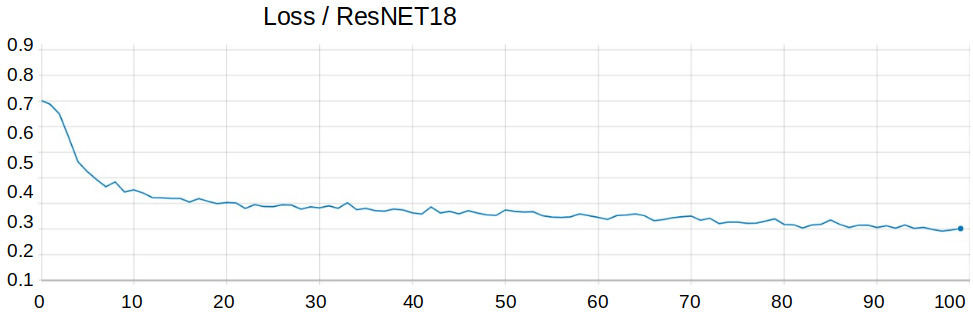}
\caption{Variation of the numerical value of Loss for the training data as a function of epoch.}
\label{fig:freq1125}
\end{figure}

Figures~\ref{fig:freq1124} and \ref{fig:freq1125} illustrate the Loss function on the
training and validating sets. We can see that
the iterative process has converged, and overfitting
is not happening. The convergence of training canalso be observed in Fig.~\ref{fig:freq1122}, \ref{fig:freq1123}, demonstrating the epoch dependences of Accuracy and F1.

\begin{figure}[h!]
\onelinecaptionstrue
\setcaptionmargin{5mm}
\includegraphics[width=0.7\linewidth]{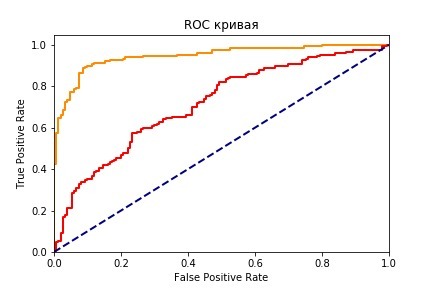}
\caption{Receiver operating characteristic (ROC). RandomForest is shown in red, and ResNet18 in orange.}
\label{fig:sim}
\end{figure}

We show that the suggested approach based
on the ResNet18 neural network model allows to
obtain better results than using the Ran-
domForest algorithm and other considered models.
In our model, the area under the ROC curve (AUC
statistics) is equal to 0.946. As mentioned in ~\cite{4} (Section 2), ROC AUC does not
depend on the choice of a single threshold. It can be shown that ROC AUC is equal to the probability of a
classifier prioritizing a randomly chosen pattern gen-
erated from the sample of objects with the SZ-effect,
higher than a randomly chosen example generated
from the non-SZ-effect sample.

We also show the receiver operating characteristic of our model. The curve is a graph that shows the ratio between the number of correctly classified objects with the pattern and the number of objects which did not carry a pattern, but were designated as correctly identified. The curve in Fig.~\ref{fig:sim} shows clearly that ResNet18 outperforms RandomForest significantly regardless of threshold.

\subsection{Comparing machine learning models with
MMF1, MMF3, and PwS using the Recall metric}

We have a set of 1000 objects with known SZ--effects. We can compare these data with simple metrics only in terms of Recall. We compared the Recall of ResNet18 and RandomForest with the earlier mentioned simple methods MMF1 and MMF3 \cite{Melin, Herranz}, PwS (PowellSnakes) \cite{PowellSnakes}. The multi-frequency coherent MMF filter increases the contrast ($\text{S}/\text{N}$ relation) of objects with known form and radiation law obtained from a series of observations, containing correlated contamination signals. Its use allows to extract SZ-effect objects using multi-frequency maps. This method is based on a universal frequency dependence of thermal radiation in objects with SZ-effect assuming that the electrons in these objects are not relativistic, and uses the spatial template (grids, metrics, etc.).

The architecture of MMF1, MMF3, and PwS is in the following \citep{rc_planck}. First a synthetic CMB map is computed for a certain model galaxy cluster profile, e.g., from \citep{2}. The map is computed for every frequency bin. The angular size of the model cluster may serve as a parameter for such a set of maps. The maps obtained from obser- vations are then convoluted with synthetic maps. The positions of peaks in such a convolution indicate the candidate positions in the observed clusters. Thus maximizing the peak values, we can use the angular size of the cluster in the model to estimate, for example, and angular sizes of the observed clusters.

Formally, the MMF1, MMF3, and PwS methods have different error minimization algorithms in cluster position and parameter computations. They are conceptually different in the fact that the PwS method is based on the Bayesian inference to assess the hypotheses (in this case, estimating the cluster parameters) (\citep{3,4}. The PSZ2 catalog from \citep{planck_zs} comprises three joint catalogs using MMF1, MMF3, and PwS methods. The PSZ2 catalog has the objects with the SZ--effect that were discovered using one method but not another. Thus, for each method the PSZ2 catalog has objects that were detected using it --- True positive (TP), and those that were not detected using it—False positive (FP). Based on these data, we can estimate and compare by Recall.

The results of the comparison of MMF1, MMF3, and PwS with RandomForest and ResNet18, presented in Table~\ref{table:comparison}, indicate that in terms of Recall metrics, our ResNet18 model exceeds the other methods, but we should remember that data analysis methods differ and the comparison is possible only for Recall.

\subsection{Comparing the detection quality on model data}

In order to check the detection of objects using the proposed method, we used the GLESP \cite{glesp1} package to generate random Gaussian maps with various $\text{S}/\text{N}$ ratios for objects with SZ-effect. We used the following coefficients in our SZ-signal approximations to construct the models: $k_{100}=-0.4$, $k_{100}=-0.4$, $k_{143}=-0.5$, $k_{217}=0.0$, $k_{353}=1.0$, $k_{545}=0.5$ for 100, 143, 217, 353, and 545\,GHz correspondingly. The relative coefficients were computed from observational data on the Abell 2319 cluster \cite{Abell, Planck1, Planck2}. The model source was generated with a $5\arcmin$ FWHM. The $\text{S}/\text{N}$ ratio was computed for a signal at 353\,GHz and scaled proportionally at other HFI frequencies.

\begin{figure}[h!]
\onelinecaptionstrue
\setcaptionmargin{5mm}
\includegraphics[width=0.7\linewidth]{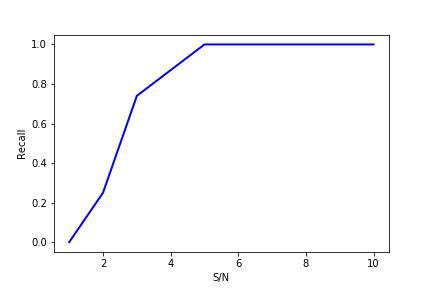}
\caption{Quality (Recall) of effect detection based on model data for different S/N ratios.}
\label{fig:sim1}
\end{figure}

In total we generated 100 radio source models. In all generated objects the SZ-effect should be observed. Fig.~\ref{fig:sim1} shows the results of comparing the quality (Recall) for different $\text{S}/\text{N}$ ratios

As shown in Fig.~\ref{fig:sim1}, the response quality of the model (Recall) increases as the task simplifies. For $\rm S/N = 1$ the network predicts that there is no effect in all objects. As the signal grows, the network confidence increases, and at $\rm S/N = 5$ and higher, the network classifies all objects correctly. Using cubic interpolation, we obtained the following result: $\rm Recall = 0.873$ corresponds to $\rm S/N \approx 3.42$. Detection of SZ-effect objects with the declared accuracy in possible at approximately the same $\text{S}/\text{N}$ ratio.

\section{CONCLUSIONS}
\label{sec:conclusion}

We constructed a model for the search of objects with the SZ-effect~\footnote{\tt https://github.com/SunnyientDev/SZ-detection}. We show that using the proposed approach based on the ResNet18 neural network model allow one to achieve a better quality than in the case of the RandomForest algorithm and other considered models. We demonstrate that detecting SZ-effect objects is possible with the declared quality $\rm Recall = 0.873$ for a ratio of $\rm S/N \approx 3.42$ and higher.

We show that this approach works and can be used for data analysis in a search for objects with the SZ--effect. The implemented approach supplements the existing algorithms \cite{Bonjean2019} of searching for such objects, allowing to work with data in GLESP package format.

{\small
{\bf ACKNOWLEDGMENTS}

The authors are grateful to the "Traektoria" foundation for scientific, cultural, and educational initiatives, and also to V. S. Ivashkin for help in this work. The work makes use of the GLESP package \footnote{\tt http://www.glesp.nbi.dk} for analyzing extended sources on a sphere. The authors are grateful to the referee for useful suggestions which helped improve the text of the paper.

{\bf CONFLICT OF INTEREST}

The authors declare no conflict of interest.}
\selectlanguage{english}

\section{APPENDIX}
\label{sec:conclusion1}

This work is based on metrics that work with two classes. Let us call the objects that have the SZ-effect the positive class ($P$), and the objects without the effect as negative ($N$). We shall refer to the responses that we know beforehand as ground truth (gt), and the responses that we predict, as prediction (pred).

Let us first consider one object of a sample. We know the correct answer beforehand (P or N) and we predict a certain response ($P$ or $N$). There are four options:

\begin{itemize}
    \item $gt = P$, $pred = P$ --- here the SZ-effect reallywas present, and we predicted it. We shall call this true positive ($TP$);
    \item $gt = N$, $pred = N$ --- there was no SZ-effect, and we predicted its absence. This shall be ``true negative'' ($TN$);
    \item $gt = P$, $pred = N$ --- SZ-effect was present, but we did not notice it. This is a ``false negative'' ($FN$);
    \item$ gt = N$, $pred = P$ --- no SZ-effect, but we think it was present. This is a ``false positive''($FP$).
\end{itemize}

For the list of sample objects we have a list of correct responses (e.g., $gt=PPPNNNN$) and a list of our predictions (e.g., $pred=PNNPPPN$)). Each object is related to one of the four cases: $TP$, $TN$, $FP$, $FN$.

Based on these principles, we can formulate all metrics which are possible for the two classes. We are interested in two:
\begin{itemize}
    \item  Accuracy --- the ratio of the fraction of objects with correctly identified SZ-effect to the total objects in the sample. i.e.,
    \begin{equation*}
    \frac{TP+TN}{TP+TN+FP+FN}
    \end{equation*}
    \item Recall --- the number of objects with $gt = P$ predicted as $P$. i.e.
    \begin{equation*}
    \frac{TP}{TP+FN}
    \end{equation*}
    
    \begin{equation*}
    \frac{TP}{TP+FP}
    \end{equation*}
    \item F1 --- the harmonic mean between Precision and Recall, i.e.,
    \begin{equation*}
    \frac{2\cdot TP}{2\cdot TP+FP+FN}
    \end{equation*}
    \item ROC AUC --- the area below the ROC curve. i.e.,
    \begin{equation*}
    \frac{1+ TPR - FPR}{2}
    \end{equation*}
    
    where True Positive Rate (TPR) is the percentage of class 1 points which were classified correctly by our algorithm, and False Positive Rate (FPR) is the percentage of class 0 points which were identified incorrectly by our algorithm.

    \end{itemize}
The Accuracy metrics can be demonstrated using the following example:

\begin{itemize}
  \item  $gt = PPPPNNNN$, $pred = NNNNNNNN$ and Accuracy $= 0.5$;
  \item $gt = PPPPNNNN$, $pred = PNPNPNPN$ and Accuracy $= 0.5$;
  \item $gt = PNNNNNNN$, $pred = NNNNNNNN$ and Accuracy $= 7/8$.
\end{itemize}

As we can see, for balanced samples (when $gt$ consists of roughly the same number of $P$ and $N$), Accuracy gives a result no lower than $0.5$, for a constant prediction of always $N$ or for a random value. But for the non-balanced classes it can give a higher value, which in this case shows not the quality of our predictions, but the balance of our sample. Result: Accuracy is useful only in balanced samples. 

The limitations of using the Recall metric can be demonstrated, for examples, as:
\begin{itemize}
  \item $gt = PPPPNNNN$, $pred = NNNNNNNN$ and Recall $= 0$;
  \item $gt = PPPPNNNN$, $pred=PNPNPNPN$ and Recall $= 0.5$;
  \item $gt = PPPPNNNN$, $pred=PPPPPPPP$ and Recall $= 1$;
  \item $gt = PNNNNNNN$, $pred=PPPPPPPP$ and Recall $= 1$.
\end{itemize}

Obtaining $1$ in Recall is not very hard—one must only always predict $P$. Therefore, the Recall metric cannot be considered separately from the others: one may not notice that the model simply yields $P$ instead of smart predictions.

The metrics work according to the following principle:
\begin{itemize}
\item The sample is balanced, therefore we can use Accuracy.
\item If we compare with the catalogs, the result is that the catalogs contain only the correct objects (TP), and the quality of the catalog can be measured only by the absence of some objects (FN). Accuracy also requires TN and FP, which we do not have. We can therefore only use Recall.
\item We can calculate the Recall for the MMF1, MMF3, and PwS catalogs, since they were compiled based on Recall maximization.
\end{itemize}

Our neural network gives a number from 0 to 1 as a result: the probability that an object belongs to a class, and we need an answer of P or N. We therefore need to impose a certain threshold, above which we get P. We divide the dataset into three parts:

\begin{itemize}
\item Train --- we learn, i.e., shift the network weights in such a way as to minimize the loss function;
\item Validation ---subsample used for setting the threshold, i.e., thresholds from 0 to 1 are examined and the one that gives a better average Accuracy for the validation set is selected;
\item Test --- this is the subsample used to measure the final Accuracy and Recall, which are then used in the table for comparison.
\end{itemize}


\begin{thebibliography}{99}

\bibitem{zs}
Ya.~B.~Zeldovich and R.~A.~Sunyaev, Astrophys.~Sp. Sci. {\bf 4}, 301 (1969).

\bibitem{Allen} S.~W.~Allen, A. E. Evrard, and A. B. Mantz. Annual Review of Astronomy and Astrophysics, {\bf 49}, 409 (2011).

\bibitem{Kravtsov} A.~V.~Kravtsov and S.~Borgani. Annual Review of Astronomy and Astrophysics, {\bf 50}, 353 (2012).

\bibitem{Nagai} D.~Nagai. In American Institute of Physics
Conference Series, {\bf 1632}, 88 (2014).

\bibitem{Sarazin} C.~L.~Sarazin. Reviews of Modern Physics, {\bf 58}, 1 (1986).

\bibitem{Kormendy} J.~Kormendy and S.~Djorgovski.Annual Review of Astronomy and Astrophysics, {\bf 27}, 235 (1989).


\bibitem{Bykov} A.~M.~Bykov, H.~Bloemen, and Y.~A.~Uvarov. Astronomy and Astrophysics, {\bf 362}, 886 (2000).

\bibitem{Basu} K.~Basu, J.~Erler, J.~Chluba et al. Bulletin of the American Astronomical Society, {\bf 51}, 302 (2019).

\bibitem{planck_zs}
Planck Collaboration: P.~A.~R.~Ade, N.~Aghanim, M.~Arnaud, M. et al., Astronomy and Astrophysics, {\bf 594}, 19 (2016).

\bibitem{Barbosa1996}
D.~Barbosa, J.~G.~Bartlett, A.~Blanchard, et al., Astron. Astroph., {\bf 314}, 13 (1996).

\bibitem{2009MNRAS.397L..41C}
D.~Cunnama, A.~Faltenbacher, C.~Cress, et al., MNRAS, {\bf 397}, L41 (2009).

\bibitem{dsol_model}
 D.~I.~Solovyov, O.~V.~Verkhodanov., Astrophys.~Bull., {\bf 72}, 217 (2017).
 
\bibitem{spt_zs}
K. Vanderlinde et al., ApJ, {\bf 722}, 1180 (2010). 
The Astrophysical Journal, Volume 722, Issue 2, pp. 1180-1196 (2010).

\bibitem{act_zs}
M.Hasselfield, M.Hilton, T.A.Marriage., JCAP, {\bf 07}, 008 (2013).

\bibitem{Abell}
G.~O. Abell, H.~G. Corwin and  R.~P. Olowin., Astrophysical Journal Supplement, {\bf 70}, 1 (1989).


\bibitem{Planck2}
Planck Collaboration, Planck 2018 results, VI. Cosmological parameters, Astronomy and Astrophysics, {\bf 641}, 67 (2020). 


\bibitem{Planck1}
Planck Collaboration, Planck 2018 results, I. Overview and the cosmological legacy of
Planck, Astronomy and Astrophysics, {\bf 641}, 56 (2020).


\bibitem{rg_list1}
M.~L.~Khabibullina and O.~V.~Verkhodanov, Astrophys. Bull. {\bf 64}, 123 (2009).

\bibitem{rg_book}
O.~V.~Verkhodanov et al.(2009).

\bibitem{blum_miley} G.~Blumenthal and G.~Miley, \aaa {\bf 80}, 13 (1979).

\bibitem{par_big3a}
 Yu.~N.~Parijskij,  W.~M.~Goss,  A.~I.~Kopylov et al., Bull. SAO, {\bf 40}, 5 (1996).

\bibitem{uss_list} C.~de~Breuck,   W.~van~Breugel,  H.~J.~A.~R\"ottgering and G.~Miley, \aas, {\bf 143}, 303 (2000).

\bibitem{rg_protocl}
B.~P.~Venemans, H.~J.~A.~Rottgering, G.~K.~Miley,  et al., Astron. Astrophys.,  {\bf 461}, 823 (2007).

\bibitem{tim_clust}
T.~V.~Keshelava and O.~V.~Verkhodanov., Astrophys. Bull., {\bf 70}, 257 (2015).

\bibitem{rc_planck}
O.~V.~Verkhodanov, E.~K.~Majorova, O.~P.~Zhelenkova, et al., Astrophys. Bull., {\bf 70}, 156 (2015).

\bibitem{wenss_sz}
O.~V.~Verkhodanov, N.~V.~Verkhodanova, O.~S.~Ulakhovich, et al., Astrophys. Bull., {\bf 73}, 1 (2018).

\bibitem{traj_zaporozhets}
A.~A.~Zaporozhets, O.~V.~Verkhodanov, Astrophys. Bull. {\bf 74}, 247 (2019).

\bibitem{oron_habr}
A.~Oronovskaya, {\tt https://habr.com/ru/users/sunny\_space/posts/}, (2018).

\bibitem{Dobbels} 
W.~Dobbels, M.~Baes, S.~Viaene, et al., Astronomy and Astrophysics, {\bf 634}, 23 (2020).

\bibitem{Tacchella}
S.~Tacchella, B.~Diemer, L.~Hernquist, et al., MNRAS {\bf 487}, 5416 (2019).

\bibitem{Ucci}
G.~Ucci, A.~Ferrara, S.~Gallerani, et al., MNRAS {\bf 465}, 1144 (2017).

\bibitem{Baron}
D.~Baron, D.~Poznanski, MNRAS {\bf 465}, 4530 (2017).

\bibitem{Aniyan}
A.~K.~Aniyan, K.~Thorat, ApJS {\bf 230}, id. 20 (2017).

\bibitem{Tao}
Y.~Tao, Y.~Zhang, C.~Cui, et al., Astr. Data Analysis Soft. and Syst. XXVII, {\bf 523}, (2018).

\bibitem{Bonjean2019}
V.~Bonjean, Astronomy and Astrophysics, {\bf 634}, 11 (2020).


\bibitem{Melin} 
J.~BMelin, J.~G.~Bartlett and J.~Delabrouille, A\&A, {\bf 459}, 3411 (2006).

\bibitem{Herranz} D.~Herranz, J.~L.~Sanz, M.~P.~Hobson et al., MNRAS, {\bf 336}, 1057 (2002).

\bibitem{PowellSnakes}
P.~Carvalho, G.~Rocha,  M.~P.~Hobson, A.~Lasenby, MNRAS, {\bf 427}, 1384 (2012).

\bibitem{glesp1}
A.G.Doroshkevich, P.D.Naselsky, O.V.Verkhodanov, et al., Int. J. Mod. Phys. D {\bf 14}, astro-ph/0305537, 275 (2003).


\bibitem{Hossin2015}
M.~Hossin, M.N.~Sulaiman, Internat. J. of Data Mining and Knowledge Manag. Proc., {\bf 5}, 1 (2015).

\bibitem{glesp3}
A.~G.~Doroshkevich, O.~V.~Verkhodanov, P.~D.~Naselsky, et al.,{Int. J. Mod. Phys.} D {\bf 20}, 1053 (2011).

\bibitem{AlexNET1}
A.~Krizhevsky, I.~Sutskever, G.~E.~Hinton, In Advances in neural information processing systems, 1097 (2012).

\bibitem{VGG1}
K.~Simonyan and A.~Zisserman, arXiv preprint, arXiv:1409.1556, (2014).

\bibitem{Hochreiter}
S.~Hochreiter, Y.~Bengio, P.~Frasconit et al., In A Field Guide to Dynamical Recurrent Neural Networks , edited by S. C. Kremer and J. F. Kolen . : IEEE Press, (2001).

\bibitem{resNET1}
K.~He, X.~Zhang, S.~Ren, S., et al., arXiv e-prints, arXiv:1512.03385, 770 (2015).

\bibitem{InceptionNet1}
C.~Szegedy, W.~Liu, Y.~Jia, P.~Sermanet et al., IEEE Conf. on Comp. Vision and Pattern Recog., INSPEC Accession Number: 15523970, DOI: 10.1109/CVPR.2015.7298594, (2015).

\bibitem{MobileNet}
A.~G.~Howard, M.~Zhu, B.~Chen et al., arXiv preprint arXiv:1704.04861, (2017).


\bibitem{NEURIPS2019_9015}
A.~Paszke, S.~Gross, F.~Massa, et al.,  arXiv:1912.01703, (2019). 

\bibitem{Russakovsky}
O. Russakovsky, J. Deng, H. Su et al., arXiv e-prints, arXiv:1409.0575, (2014).

\bibitem{liu2019radam}
L.~Liu, H.~Jiang, P.~He, W.~Chen et al., Published as a conf. paper at ICLR, arXiv:1908.03265, (2020).

\bibitem{Fmera} D.~M.~W.~Powers, School of Informatics and Engineering
Flinders University, Australia Technical Report SIE-07-001, web.archive.org, (2007). 

\bibitem{2} M.~Arnaud, G.~W.~Pratt, R.~Piffaretti et al.,  
Astronomy and Astrophysics, {\bf 517}, 20 (2010). 

\bibitem{3} P.~Carvalho, G.~Rocha, M.~P.~Hobson, MNRAS, {\bf 393}, 681 (2009). 

\bibitem{4} P.~Carvalho, G.~Rocha, M.~P.~Hobson, A.~Lasenby, MNRAS, {\bf 427}, 1384 (2012).  

\bibitem{Rengelink} R.~B. Rengelink, Y. Tang,  A.~G. de Bruyn et al., Astron. and Astroph. Sup. {\bf 124}, 259 (1997).

\bibitem{Condon} J.~J. Condon, W.~D. Cotton, E.~W. Greisen et al., The Astronomical Journal {\bf 115}, 1693 (1998).
 
 
\end{thebibliography}
\end{document}